\documentclass[conference]{IEEEtran}
\IEEEoverridecommandlockouts
\usepackage{soul}
\usepackage{todonotes}

\usepackage{cite}
\usepackage{tabto}
\usepackage{amsmath,amssymb,amsfonts}
\usepackage{algorithmic}
\usepackage{graphicx}
\usepackage{textcomp}
\usepackage{xcolor}
\usepackage{multirow}
\usepackage{balance}
\usepackage{tikz}
\usepackage{color}
\usepackage{mathtools, cuted}
\usepackage[utf8]{inputenc}
\usepackage{caption}
\usepackage{subcaption}
\def\BibTeX{{\rm B\kern-.05em{\sc i\kern-.025em b}\kern-.08em
    T\kern-.1667em\lower.7ex\hbox{E}\kern-.125emX}}

\begin{document}

\title{Data Augmentation for Histopathological Images Based on Gaussian-Laplacian Pyramid Blending%\title{Gaussian-Laplacian Pyramid Blending based Data Augmentation for Histopathological Images\\
}

\author{\IEEEauthorblockN{Steve Tsham Mpinda Ataky\IEEEauthorrefmark{1}, Jonathan de Matos\IEEEauthorrefmark{1}\IEEEauthorrefmark{4}, Alceu de S. Britto Jr.\IEEEauthorrefmark{2}\IEEEauthorrefmark{4}\\ Luiz E. S. Oliveira\IEEEauthorrefmark{3} and Alessandro L. Koerich\IEEEauthorrefmark{1}}\\ \IEEEauthorblockA{\IEEEauthorrefmark{1}École de Technologie Superiéure, Université du Québec, Montréal, QC, Canada\\
Email: steve.ataky@nca.ufma.br, alessandro.koerich@etsmtl.ca}
\IEEEauthorblockA{\IEEEauthorrefmark{4}State University of Ponta Grossa (UEPG), Ponta Grossa, PR, Brazil\\ Email: jonathan@uepg.br}
\IEEEauthorblockA{\IEEEauthorrefmark{2}Pontifical Catholic University of Paraná, Curitiba, PR, Brazil\\
Email: alceu@ppgia.pucpr.br}
\IEEEauthorblockA{\IEEEauthorrefmark{3}Federal University of Paraná, Curitiba, PR, Brazil\\ Email: luiz.oliveira@ufpr.br}
}

\maketitle

%\ninept
%
\maketitle
\begin{abstract}
Data imbalance is a major problem that affects several machine learning (ML) algorithms. Such a problem is troublesome because most of the ML algorithms attempt to optimize a loss function that does not take into account the data imbalance. Accordingly, the ML algorithm simply generates a trivial model that is biased toward predicting the most frequent class in the training data. %DA techniques have been used to mitigate the data imbalance problem. However,
In the case of histopathologic images (HIs), both low-level and high-level data augmentation (DA) techniques still present performance issues when applied in the presence of inter-patient variability; whence the model tends to learn color representations, which is related to the staining process. In this paper, we propose a novel approach capable of not only augmenting HI dataset but also distributing the inter-patient variability by means of image blending using the Gaussian-Laplacian pyramid. The proposed approach consists of finding the Gaussian pyramids of two images of different patients and finding the Laplacian pyramids thereof. Afterwards, the left-half side %of one image
and the right-half side of
%another
different HIs are joined in each level of the Laplacian pyramid, and from the joint pyramids, the original image is reconstructed. This composition %, resulting from the blending process,
combines the stain variation of two patients, avoiding that color differences mislead the learning process. Experimental results on the BreakHis dataset have shown promising gains vis-à-vis the majority of %traditional
DA techniques presented in the literature. 
\end{abstract}


%%%%%%%%%%%%%%%%%%%%%%%%%%%%%%%%%%%%%%%%%%%%%%%%%%%%
\section{Introduction}
\label{sec:intro}
%%%%%%%%%%%%%%%%%%%%%%%%%%%%%%%%%%%%%%%%%%%%%%%%%%%%
Cancer is a global health problem and can be the greatest barrier to the long life expectancy worldwide in the 21st century \cite{cancer2018}.
%Due to its impact on life expectancy and life quality, the efforts to combat the disease are stabilizing the mortality in high-income countries.
%The aforementioned efforts are focused on reducing risk factors like smoking, overweight, physical inactivity and promote early diagnosis and treatment. 
Breast cancer is the most prevalent cancer type among women in 140 out of 184 countries 
%according as stated by
\cite{torre2017}. Its detection usually starts with self-examination and periodic mammography. These exams can identify lumps that will be examined in detail by ultrasound, computed tomography, or magnetic resonance imaging. When some characteristics that can point to a malignant tumor are detected by imaging exams, the final step is the biopsy, which is considered the gold-standard in the diagnosis process because it provides the most accurate diagnosis of the tumor type. The diagnosis procedure should be fast because some malignant tumors grow very fast and have high metastasis probability. Biopsies are a complex diagnosis tool, requiring the acquisition of material (e.g. fine-needle aspiration or open surgical biopsy), tissue treatment (slicing, staining and slide preparation) and analysis
%. The analysis requires an 
by an experienced pathologist. Such an analysis besides being time-consuming, it is subject to inter and intra-observer issues \cite{belloq}. The variance in the results can be due to the pathologist experience or the hematoxylin and eosin (H\&E) staining color differences, which may be related to stain manufacturers, storage age, and temperature.

Computer-aided diagnosis (CAD) systems can help in such an analysis by adding an extra opinion to the pathologist decision. CAD systems can rely on histopathologic images (HIs), which are obtained from tissue slides scanning, to make decisions about tumor characteristics, e.g. benign or malignant \cite{deMatos2019}. The automatic classification of HIs is a challenging problem in machine learning (ML) because HIs do not have the same structural aspect of macroscopic images such as people’s faces, cars, animals, or traffic signs. One structure that is important in HIs is the nucleus. In images stained with H\&E, hematoxylin highlights the nuclei with a blueish color, and eosin highlights the cytoplasm and extracellular matrix in pink. The importance of the nucleus in tumor diagnosis is related to its quantity and format. When a region presents a highly abnormal amount of nuclei, this can be an indication of excessive cell multiplication, meaning a strong sign of a tumor. The format of the nuclei may also represent a signal of a tumor, this is called nucleus pleomorphism. Although blue and pink are the colors expected in H\&E stained slides, it is common to face differences in intensity, saturation, and hue in the HIs. When analyzing an HI dataset, it is possible to note the differences
%, which are more evident when looking into images of different patients.
in the slides from different patients, when they have been produced using different
%used other
stain brands or faced variations during the whole process. The pathologist can understand this variation easily due to its expertise in looking at important HI features. {\color{black} However, color variations between patients may introduce a bias to ML algorithms. The image color of %one
a patient's slide does not vary %because
if the staining process is the same, but the inter-patient slide color, even for the same tumor type, may be different. It is also possible to have images of different patients with similar colors, even %with one being benign and another a malignant tumor.
if the tumor types are different.} Therefore, color normalization may help minimize the color bias. However, color normalization algorithms
%and it
require a target image as a reference to guide the normalization process \cite{norm01, norm02}, which is a complex task that may cause loss of important characteristics. Furthermore, the classification result is closely related to such a reference image.

Another aspect related to the classification of HIs is the number of images available to train an ML model, especially
%One fact that made possible the latest evolution in ML is the development of
deep learning (DL) models. DL models have been achieving impressive performance
%mainly in daily life problems related to images,
in several image classification tasks. However, as DL models usually have millions of parameters to tune, they require very large datasets for training to avoid overfitting the models. 
%Low data scenarios are not adequate for DL because it produces large models with millions of parameters, which leads to overfitting problems. 
HI datasets, on the other hand, are usually small because analyzing and labeling HIs is expensive and requires experienced pathologists. Data augmentation (DA) has been actively used to circumvent this issue. Besides, HI datasets can also suffer from the data imbalance due to tumor occurrence rate and the biopsy priorities. As the biopsy procedure is expensive and time-consuming, it is usually carried out when malignant tumors are previously diagnosed through non-invasive exams.%On the other hand,
Furthermore, the biopsy is not required when most of the benign tumor types are detected. Therefore, it is common in the case of breast cancer biopsies to have a higher number of malignant HIs than benign HIs. Such a data imbalance also impacts on ML algorithms.%, but there are some approaches that attempt to circumvent this problem.

In this paper, we propose a novel approach for augmenting HI datasets, which distributes the inter-patient variability by means of image blending using the Gaussian-Laplacian pyramid.
{\color{black} The main contributions of this paper are: (i) a novel data augmentation method that simultaneously provides color normalization and augmentation; (ii) a method to improve data classification by reducing the data imbalance; (iii) the advantages of using a texture convolutional neural network for HI classification.} This paper is organized as follows: Section \ref{sec:relwork} presents related works found in the literature to data augmentation and data balancing, as well as their advantages and limitations. Section \ref{sec:method} briefly describes Gaussian-Laplacian pyramids, pyramid blending, and presents the proposed approach for data augmentation. Section \ref{sec:experiments} describes the experimental setup used in this work. Results and conclusions are reported in Sections \ref{sec:results} and \ref{sec:conclusion}, respectively.

% ALEKOE OK.
%%%%%%%%%%%%%%%%%%%%%%%%%%%%%%%%%%%%%%%%%%%%%%%%%%%%
\section{Related Works} \label{sec:relwork}
%%%%%%%%%%%%%%%%%%%%%%%%%%%%%%%%%%%%%%%%%%%%%%%%%%%%
One way to circumvent the overfitting problem due to the small and imbalanced datasets is through DA. DA strategies have been exploited in many articles related to HIs, usually applying low-level transformation such as flip and rotation \cite{dataauggeo01, dataauggeo02, dataauggeo03, dataauggeo04}. %Notwithstanding, based on the principle that transformations are motion geometry, the shape/object created during a transformation is an image of the original one.
Rotation, which consists of a turn of an image makes a congruent image of the original but facing another direction. Flip, likewise, creates a congruent mirror image of the original one. In both cases, the number of images increases but the heterogeneity in terms of texture is preserved. In other words, %considering HIs,
low-level transformations solve overfitting solely in terms of the number of images, but not in terms of inter-patient variability. Patching is another DA strategy that has been used with HIs~\cite{patchbreakhis, alexbreakhis, MatosBOK19}. This DA strategy divides an HI into some patches (overlapped or not). Thus, from one sample \textit{n} samples are generated. Notwithstanding the gain in terms of the amount of data, some patches may not contain meaningful information in light of their size, the magnification of the original HI as well as their location in the original HIs. In the same way, some works also used DA strategies based on color disturbance~\cite{colordataaug02}. Color disturbance contributed not only to increase the number of images but also to eliminate the color bias.

\textcolor{black}{In addition to
%geometric transformations, color space augmentations, random cropping, feature space augmentation,
such DA strategies, generative adversarial networks (GANs)
%stands as one of the most used in Computer Vision, meanly for training DL models. GANs modeling
have also been used to create artificial instances from a particular dataset in such a way that the generated images retain similar characteristics vis-à-vis the original ones~\cite{gans1}. Besides, GANs have capabilities to mimic data distributions as well as to synthesize input images at remarkable levels of realism, because they maximize the probability density over the data by exploiting density ratio estimation~\cite{gans2}. Furthermore, GANs can find out the high dimensional latent distribution of data, which is the reason for the significant performance gains in terms of visual feature extraction. GANs have been used in medical imaging
%, different modalities, to wit, Computed Tomography, chest X-Ray, Ultrasound, Magnetic Resonance Imaging, and Microscopy used methods based on GANs for
not only for DA but also for de-noising, segmentation, and reconstruction~\cite{gans3}}

Concerning the data imbalance, due to tumor occurrence rate and the biopsy priorities, several approaches have been proposed to remedy this problem such as cost functions, ensemble learning, as well as algorithm-level and data-level approaches~\cite{imbalance}. Data-level approaches generate new samples to balance the dataset. In the literature, it is also found some key techniques that are used to balance data at feature-level, such as SMOTE~\cite{dataaugfeat1}, ADASYN~\cite{dataaugfeat2}, and ROSE~\cite{dataaugfeat3}. These techniques introduce synthetic examples through the interpolation between various positive instances that lie together.
%Though balancing methods with the proposed one converge in terms of balancing purpose, the latter is suitably employed but while using shallow classifiers, such as support vector machines, k-nearest neighbors, owing to the fact that those are focused on the feature space and these learn from extracted features. Moreover, the resulting samples on the original feature space are limited to the local space given that the interpolation of examples of the less frequent classes %of the synthetic between the smaller class example processing and one of its nearest neighbors. When the neighbor is far from the center, which means that there are merely a small number of examples in the local space near the center, and the true underlying distribution of the class will simply be expressed unreliable.
In the case of HIs, this means that new samples will be the approximation of existing samples, not solving the inter-patient diversity nonetheless.
% ALEKOE OK

%%%%%%%%%%%%%%%%%%%%%%%%%%%%%%%%%%%%%%%%%%%%%%%%%%%%
\section{Proposed Method} \label{sec:method}
%%%%%%%%%%%%%%%%%%%%%%%%%%%%%%%%%%%%%%%%%%%%%%%%%%%%
The proposed method aims to improve the generalization ability of ML algorithms dealing with HIs, by considering the inter-patient variability. We propose the use of a blending method to composite images of different patients with the same type of tumor (benign or malignant). The present approach generates a new training image made up of half images of different patients. This strategy aims to avoid that a model learns color representations of patients, which, in fact, are related but to the staining process. While the straightforward blending of two images may produce artifacts owing to the adjacent pixel intensity difference, to remove such artifacts Gaussian-Laplacian pyramid was used.

\begin{figure*}[htp!]
\begin{minipage}[b]{1.0\linewidth}
  \centering
  \centerline{\includegraphics[width=4.0cm]{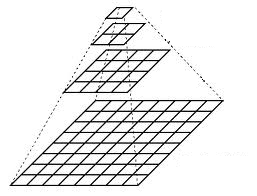}}
  \centerline{(a) Pyramid representation}\medskip
\end{minipage}
\begin{minipage}[b]{.48\linewidth}
  \centering
  \centerline{\includegraphics[width=5.0cm]{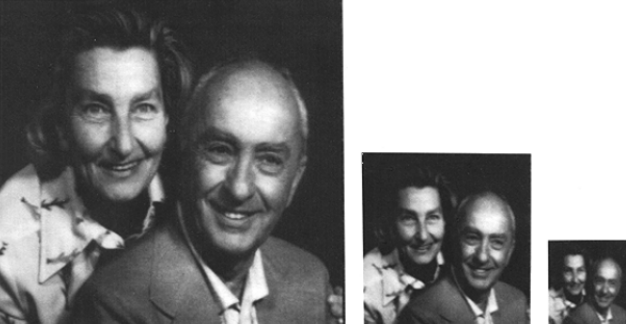}}
  \centerline{(b) Gaussian pyramid}\medskip
\end{minipage}
\hfill
\begin{minipage}[b]{0.48\linewidth}
  \centering
  \centerline{\includegraphics[width=5.0cm]{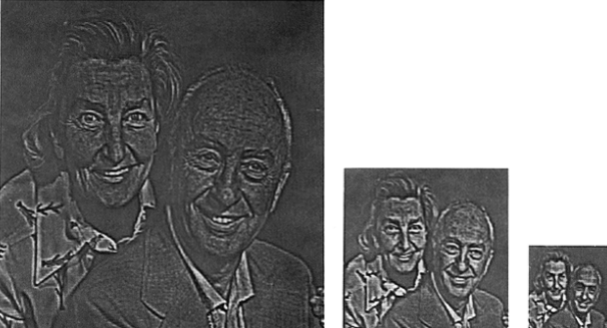}}
  \centerline{(c) Laplacian pyramid}\medskip
\end{minipage}
\caption{An example of Gaussian and Laplacian pyramids from the same input image. (b) First three levels of Gaussian pyramid; (c) First three levels of Laplacian pyramid. Images adapted from \cite{adelson1984pyramid}.}
\label{fig:pyramids}
\end{figure*}

%%%%%%%%%%%%%%%%%%%%%%%%%%%%%%%%%%%%%%%%%%%%%%%%%%%%
\subsection{Gaussian-Laplacian Pyramid (GLP)}
\label{sec:GLP}
%%%%%%%%%%%%%%%%%%%%%%%%%%%%%%%%%%%%%%%%%%%%%%%%%%%%
In signal processing and computer vision, pyramid representation (Fig.~\ref{fig:pyramids}a) is the main type of multi-scale representation for computing image features on different scales. The pyramid is obtained by repeated smoothing and subsampling of an image or a signal. This concept is frequently used because it expresses computational efficiency approximation compared to other representations such as scale-space representation and multi-resolution analysis \cite{crowley2003fast, lowe2004distinctive,burt1981fast}. For generating the pyramid representation, different smoothing kernels have been brought forward and the binomial one strikingly shows up as useful and theoretically well-founded \cite{lindeberg2013scale}.

Accordingly, for a bi-dimensional image, the normalized binomial filter may be applied (1/4, 1/2, 1/4) in most cases twice or even more along all spatial dimensions, afterward, the subsampling of the image by a factor of two, which leads to efficient and compact multi-level representation. There are two main types of pyramids, namely, low-pass and band-pass \cite{burt1983laplacian, adelson1984pyramid}. In order to develop filter-based representations by decomposing images into information on multiple scales as well as to extract features/structures of interest from an image, Gaussian pyramid (GP), Laplacian pyramid (LP), and wavelet pyramid are examples of the most frequently used pyramids.

{\color{black} The GP consists of low-pass filtered, reduced density,} where subsequent images of the preceding level of the pyramid are weighted down by means of Gaussian average (or Gaussian blur) and scaled down, as shown in Figs.~\ref{fig:pyramids}a and~\ref{fig:pyramids}b. The base level ($l=0$) is defined as the original image. Formally speaking, assuming that $I(x,y)$ is a two-dimensional image, the GP is recursively defined in \eqref{eq:GP}.

\begin{strip}
\begin{equation}
	\begin{aligned}
		G_l(x,y) =\begin{cases}
		    I(x,y), & \text{for level } l=0\\
		    \\
			\sum\limits_{m = -2}^{2} \sum\limits_{n = -2}^{2} w(m, n)G_{l-1}(2x+m, 2y+n), & \text{otherwise.}
		\end{cases}
	\end{aligned}
	\label{eq:GP}
\end{equation}
\end{strip}

%\todo[inline]{ALEKOE -- $w(m, n)$ in Eq.1, not $w(x, y)$ right? $m$ and $n$ were not defined as well as $x$ and $y$.}

\noindent where $m$ and $n$ are pixel coordinates and $w(m, n)$ is a Gaussian kernel %weighting function (identical at all levels) termed the generating kernel
which adheres to the following properties: separable, symmetric and each node at level $l$
%$n$
contributes the same total weight to nodes at level $l+1$. The pyramid name arose from the fact that the Gaussian kernel nearly approximates a Gaussian function. This pyramid holds local averages on different scales, which has been leveraged for target localization and texture analysis \cite{burt1983fast, larkin1983multi, anderson1985change}.

{\color{black} In order to seamlessly stitching together image $I_A$ and image $I_B$ into a composite image on a scale-dependent way such that to avoid boundary artifacts, the LP is used. LP uses GP to blend images by preserving the significant feature meanwhile, as shown in Fig.~\ref{fig:pyramids}c. The process is performed by downsizing the images into different levels (sizes) with Gaussian. Afterward, the Gaussian is expanded into the lower level and subtracts from the image at that level to acquire the Laplacian image. In other words, a level in LP is formed by the difference between that level in GP and expanded version of its upper level in GP. The smallest level, however, is not a different image for enabling the high-resolution image reconstruction.}
%\todo[inline]{ALEKOE -- Something inconsistent here. What is $[\vec{I}_{0}, \vec{I}_{1}, \dots, \vec{I}_{n}]$?? $I(x,y)$ was defined as an Image. GP for me should be $G_0, G_1, \dots G_?$ not $G_n$ because $n$ was already used. Why using $[\vec{I}$ if ${I}$ is an image, not a vector?}
Formally speaking, assuming the GP $[{G}_{0}, {G}_{1},\dots, {G}_{k}]$, where $k$ denotes the number of levels,  %$[{I}_{0}, {I}_{1}, \dots, {I}_{k}]$,
the LP is obtained by computing:

\begin{strip}
\begin{equation}
	\begin{aligned}
		L_l(x,y) =	G_{l}(x,y)-4\sum\limits_{m=-2}^{2} \sum\limits_{n=-2}^{2} w(m, n)G_{l+1} \left(\frac{x-m}{2}, \frac{y-n}{2} \right)
	\end{aligned}
	\label{eq:LP}
\end{equation}
\end{strip}

\noindent where $L_l(x,y)$ is the difference between $G_{l}(x,y)$ and an upsampled, smoothed version of $G_{l+1}(x,y)$.
%$	L_l(x,y)={G}_{l}(x,y)-E({G}_{l+1}(x,y))$ %${b}_{k}={I}_{k}-E{I}_{k+1}$, where $E{I}_{k+1}$ represents an upsampled, smoothed version of ${I}_{k+1}$ of the same dimension.
In the literature, LP is used for image compression, image enhancement, image analysis, and graphics \cite{adelson1984pyramid}.

%%%%%%%%%%%%%%%%%%%%%%%%%%%%%%%%%%%%%%%%%%%%%%%%%%%%
\subsection{Pyramid Blending (PB)}
\label{sec:blending}
\begin{figure}[htpb!]
     \centering
     \begin{subfigure}[b]{0.23\textwidth}
         \centering
         \includegraphics[width=3.5cm]{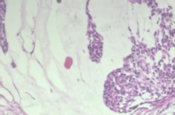}
         \caption{$I_A$}
         \label{fig:leftimg}
     \end{subfigure}
     \hfill
     \begin{subfigure}[b]{0.23\textwidth}
         \centering
         \includegraphics[width=3.5cm]{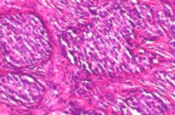}
         \caption{$I_B$}
         \label{fig:rightimg}
     \end{subfigure}
     \\
     \begin{subfigure}[b]{0.23\textwidth}
         \centering
         \includegraphics[width=3.5cm]{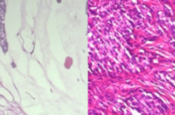}
         \caption{ $I_{C, 1}$}
         \label{fig:simpleblending}
     \end{subfigure}
     \hfill
     \begin{subfigure}[b]{0.23\textwidth}
         \centering
         \includegraphics[width=3.5cm]{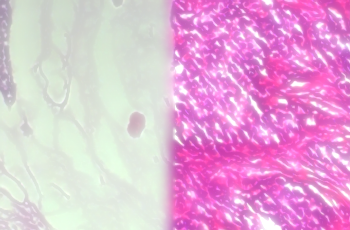}
         \caption{ $I_{C, 2}$}
         \label{fig:blendingglpb}
     \end{subfigure}
        \caption{Examples of image blending: (a) $I_A$ and (b) $I_B$ are two different images to be blended; (c) $I_{C, 1}$ is a direct blending of $I_A$ and $I_B$; (d) $I_{C, 2}$ is a multi-resolution blending of $I_A$ and $I_B$.}
        \label{fig:blending}
\end{figure}

%%%%%%%%%%%%%%%%%%%%%%%%%%%%%%%%%%%%%%%%%%%%%%%%%%%%
Blending is a common task in several scientific applications whose purpose is to join smoothly two images or objects into a larger composite image in such a way that their respective boundaries junctions are unnoticed~\cite{adelson1984pyramid}. Fig.~\ref{fig:blending} shows an example of blending images $I_A$ and $I_B$ into composite images $I_{C,1}$ and $I_{C,2}$. 

Let $I_A$ and $I_B$ be two images of the same resolution, and $I_C$ be the composite image from blending $I_{A,\text{left}}$ and $I_{B,\text{right}}$ which are the half-left side and the half-right side of $I_A$ and $I_B$, respectively. The direct blending is given by \eqref{eq:blen}. 

\begin{equation}
    I_C = I_{A,\text{left}} + I_{B,\text{right}}
\label{eq:blen}
\end{equation}

As shown in Fig.~\ref{fig:blending}c, $I_{C,1}$ is often a composite in which the boundary junction is apparent when splining two images. The transition from one image to the other in case of direct blending may carry mismatch of both low and high frequencies.

%%%%%%%%%%%%%%%%%%%%%%%%%%%%%%%%%%%%%%%%%%%%%%%%%%%%
\subsection{Proposed Approach for Data Augmentation}
\label{sec:pagestyle}
%%%%%%%%%%%%%%%%%%%%%%%%%%%%%%%%%%%%%%%%%%%%%%%%%%%%
\begin{figure}[htb]
%\begin{minipage}[b]{0.99\linewidth}
  \centering
  \centerline{\includegraphics[width=7.5cm]{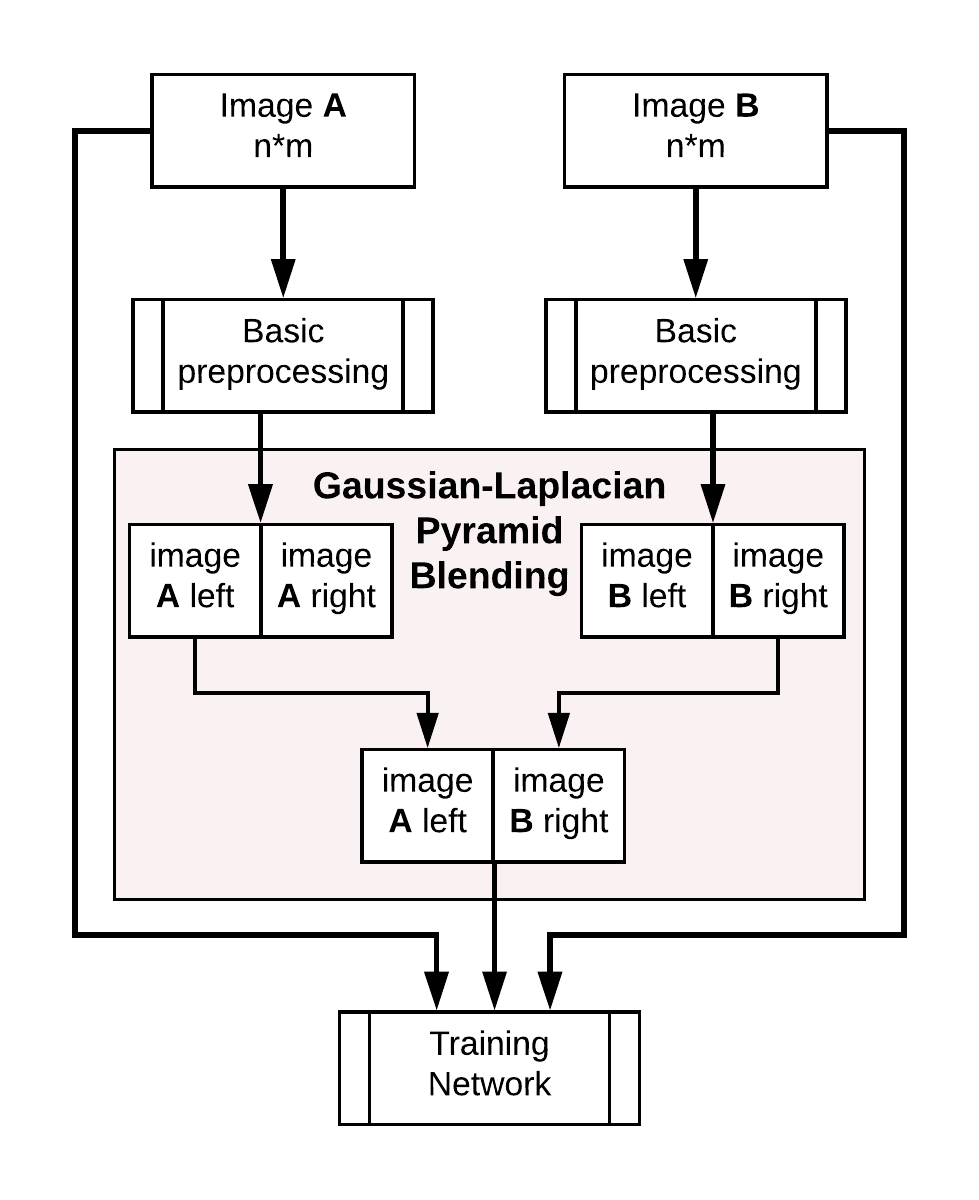}}
%  \vspace{1.5cm}
  %\centerline{(a) Results 3}\medskip
%\end{minipage}
\caption{An overview of the proposed approach for DA based on GLPB.}
\label{fig:method}
\end{figure}

%According as it is presented in the theoretical concepts (Section \ref{sec:GLP} and Section \ref{sec:blending})
This section presents an efficient approach using Gaussian-Laplacian pyramid blending (GLPB) for HI data augmentation. An overview of the proposed approach is shown in Fig.~\ref{fig:method}. The proposed approach for DA is based on blending two images of two different patients to generate the third image. As stated in Section~\ref{sec:blending}, the direct blending of $I_{A,\text{left}}$ and $I_{B,\text{right}}$ usually carries mismatch of low and high frequencies, therefore the boundary junction is evident and the resulting image does not look natural. Our hypothesis is that pyramid blending stands out as the most suitable solution to address this issue and produce natural-looking HIs. This process consists of decomposing each image into a set of spatial-frequency bands. Afterward, a band-pass composite can be constructed in each band by means of a transition zone. The latter is comparable in width to the wavelength representation in the band. To obtain the final composite, component band-pass {composites} are summed. The computational steps of the proposed multi-resolution splining procedure are quite feasible when pyramid methods are used~\cite{adelson1984pyramid}.
 
In order to stitch together {$I_{A,\text{left}}$ and $I_{B,\text{right}}$} into a composite image with minimum or no apparent junction boundaries, the Laplacian pyramid is used to smoothing the boundary on a scale-dependent way to avoid boundary artifacts. Assuming that $I_{A}$ and $I_{B}$ have the same resolution, the proposed approach is made up of the following steps:

\begin{enumerate}
    \item Input: Patient images $I_A$ and $I_B$, and a (binary) mask $R$ that specifies the blend ($0=I_A$, $1=I_B$), with $A \neq B$.
    \item Build $I_{A_i}$'s (AL) Laplacian pyramid,  $i \in 0 ... N$; $I_{B_j}$'s (BL) Laplacian pyramid, $j \in 0 ... N$, and $R_p$'s (RG) Gaussian pyramid $p \in 0 ... N$;
     \item Build a Laplacian pyramid for the result F, using
    linear interpolation over every pixel, with a blend mask given at different levels of detail, for every pyramid level k:
    \\ \qquad $FL_k(r,c) = (1-RG_k(r,c))*AL_k(r,c) + RG_k(r,c)*BL_k(r,c)$, where r and c are row and column, respectively; 
     \item Reconstitute the full-resolution image for F, by building F Gaussian (FG) and F Laplacian (FL) as follow:\\
        $F=FG_0$ from $FL_i, i \in 0 ...N$.\\
        $FG_N = FL_N $\\
        $FG_k = FL_k + EXPAND(FG_{k+1})$. 
 \end{enumerate}
 
\begin{figure*}
     \centering
     \begin{subfigure}[b]{0.49\textwidth}
         \centering
         \centerline{\includegraphics[width=8.5cm]{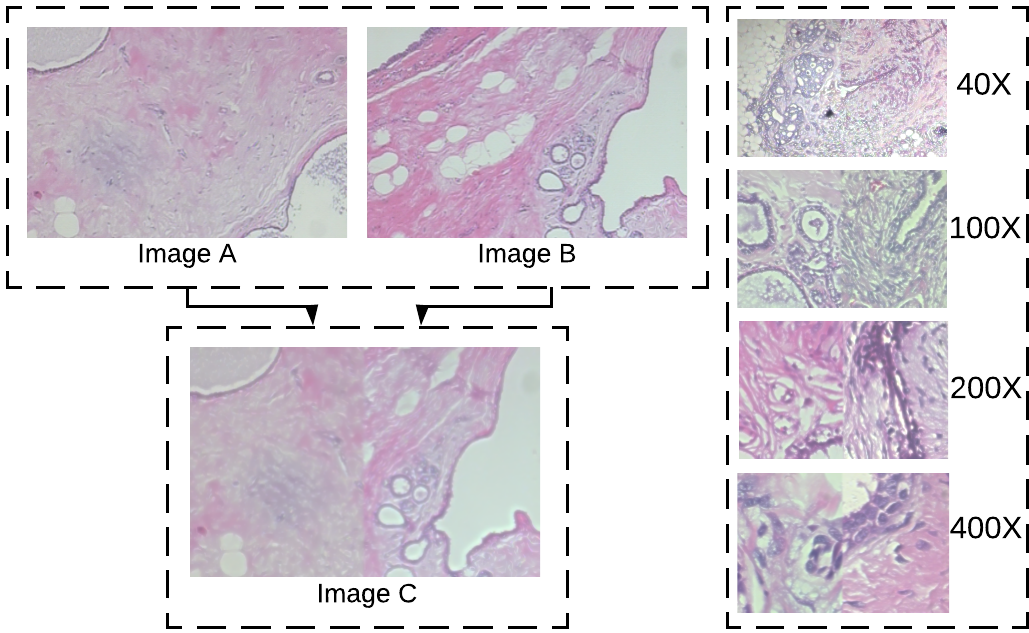}}
         \caption{Image blending with pyramids (left), examples of blending in which the stitching not evident (right).}
         \label{fig:leftimg}
     \end{subfigure}
     \hfill
     \\
     \vspace{10pt}
     \begin{subfigure}[b]{0.49\textwidth}
         \centering
         \centerline{\includegraphics[width=8.5cm]{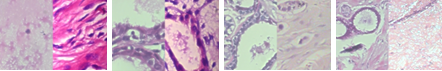}}
         \caption{Examples of blending results in which the stitching is evident.}
         \label{fig:simpleblending}
     \end{subfigure}
     \hfill
     \caption{Examples of HI blending: (a) Image C is the composite of the pyramid blending of Image A and Image B; images on the right side are composite images generated by pyramid blending of different patients. Each image is the result of different images of different magnitudes such as 40$\times$, 100$\times$, 200$\times$ and 400$\times$ as specified on the right side of each image; (b) samples in which the blending process generated artifacts.}
\label{fig:final_example}
\end{figure*}

Fig.~\ref{fig:final_example} illustrates the results achieved by the proposed method on HIs from the BreakHis dataset \cite{breakhis}.

%%%%%%%%%%%%%%%%%%%%%%%%%%%%%%%%%%%%%%%%
\section{Experimental Setup}
\label{sec:experiments}
%%%%%%%%%%%%%%%%%%%%%%%%%%%%%%%%%%%%%%%%
The dataset used in the experiments is BreakHis \cite{breakhis}. It consists of 7,909 breast cancer histopathologic images. The images are labeled following eight types of tumor, four malignant and four benign. The distribution of images is imbalanced according to the tumor type, {due to the prevalence of certain tumor types in the population}, presented in Table \ref{table:classdistribution}. The original image size is 700$\times$460 pixels and 8-bits RGB. We used images downsized by half to reduce the CNN overhead. The classification considered but benign and malignant classes, that is, all benign tumors are considered as a single class and all malignant tumors as another class. 

\begin{table}[htpb!]
\caption{Image and patient distribution in the BreakHis dataset.}
\label{table:classdistribution}
\centering
\begin{tabular}{|l|l|r|r|} 
		\hline
		\bfseries Tissue &  &  \bfseries \# of &  \bfseries \# of \\
		\bfseries Type & \bfseries Tumor Type &  \bfseries Images &  \bfseries Patients \\
		\hline
		\multirow{5}{*}{{\rotatebox[origin=c]{90}{\parbox[c]{1cm}{\centering \bfseries Benign}}}} & Adenosis & 444 & 4 \\
		& Fibroadenoma & 1,014 & 10 \\
		& Phyllodes tumor & 453 & 3 \\
		& Tubular adenoma & 569 & 7 \\
		& \multicolumn{1}{|c|}{\bf Total} & 2,368 & 24 \\
		\hline
		\multirow{5}{*}{{\rotatebox[origin=c]{90}{\parbox[c]{1cm}{\centering \bfseries Malign}}}} & Ductal carcinoma & 3,451 & 38 \\
		& Lobular carcinoma & 626 & 5 \\
		& Mucinous carcinoma & 792 & 9 \\
		& Papillary carcinoma & 560 & 6 \\
		& \multicolumn{1}{|c|}{\bf Total} & 5,429 & 58 \\
		\hline
	\end{tabular}
	\vspace{-0.1 in}
\end{table}

HIs do not have a defined geometric structure like macroscopic images do, thus, we use a compact CNN model, named Texture CNN (TCNN)~\cite{texturecnn01}. The network architecture is presented in Table~\ref{tab:tcnn} and it consists of two convolutional layers interleaved with batch normalization layers, a global average pooling, and two fully connected layers. All layers except the last one use ReLU activation function. The implementation was done using Pytorch. We used adadelta for optimization with a learning rate of 0.1 and 25 epochs. The number of images used in the experiments is reduced, taking into account that the 7,909 images are divided into four magnification factors. This is the other reason for employing TCNN, which has fewer parameters compared to other CNN architectures such as VGG or ResNet.

\begin{table}[htb]
\caption{The architecture of TCNN.}
\label{tab:tcnn}
\centering
\begin{tabular}{|r|c|c|c|}
\hline
\bf Layer & \bf Size & \bf Kernel & \bf Filters \\
\hline
Input & 350$\times$230 &  - & 3 \\
Conv2D & 350$\times$230 & 3 & 32 \\
BatchNorm & 350$\times$230 & - & - \\
ReLU & 350$\times$230 & - & -\\
Conv2D & 350$\times$230 & 3 & 32 \\
BatchNorm & 350$\times$230 & - & -\\
ReLU & 350$\times$230 & - & -\\
GlobalAveragePool & 1$\times$1 & - & 32 \\
FullyConnected & 32 & - &  -\\
ReLU & 32 & - & -\\
FullyConnected & 16 & - &  -\\
SoftMax & 2 & - &  -\\
\hline
\end{tabular}
\end{table}

We analyzed the impact of DA on TCNN by varying the augmentation factor. It is worthy of note that the comparisons herein are more focused on the imbalance problem of the dataset, meaning that we used the GLPB only to generate new images of the benign type of tumor. We have not applied the GLPB in the malignant images. We generated as many benign images as necessary to equalize the amount of benign and malignant images. The GLPB approach was evaluated using no DA (GLPB), using additional color DA over the entire set of equalized images (benign and maligned) to double the size of the training set (GLPB2$\times$) and to augment it six times (GLPB6$\times$). Therefore, all experiments with GLPB were based in the GLPB equalized dataset.

{Furthermore, we have also implemented a blending algorithm to minimize the boundary junction formed between the combination of the two images after a direct blending}. This blending (Mix) {applies a color linear combination in} the middle of the composite images (half of each image), resulting in a smooth transition between them. We applied the Mix between benign random images (MixB\&M) and we also mixed images of the same sub-type of the tumor, e.g. adenoma images of two different patients (MixSub). We also compared our results with the original dataset
%, that is, the dataset
without DA and class balancing (NoAug). Furthermore, the classification results without balancing, but using only color DA are also presented (Aug2$\times$). An interesting analysis regarding DA and TCNN can be found in \cite{texturecnn02}, which used an approach based on the color information \cite{colordataaug03}. %It applies a variable random hue, bright and contrast maximum alteration of 4\%, 25\% and 25\%, respectively. The color augmentation procedure can also work as a color normalization, reducing the bias of stain alterations. All DA procedures (blending and color DA) were applied only to the training set.
In addition, we have also used an approach based on GANs to generate a combination of images between patients. We used the same CycleGAN proposed by \cite{cyclegan}, which generates new images combining features from two origins. We trained a single model of the CycleGAN using all images from the dataset, from both the classes. During the training process, we only used pairs of the same class as samples (A and B in the CycleGAN were of the same class), so the transforming process of the GAN was restricted inside classes. The images to the augmentation process were also generated inside a single class, not mixing images from benign or malignant classes. We combined the same images used in the GLPB approach with the intention of balancing the dataset and also applied 2$\times$ (GAN2$\times$) and 6$\times$ (GAN6$\times$) DA.

In total we have carried out 10 experiments, three involving the GLPB approach, three with the GAN balanced dataset, two with no balancing and {two using the Mix approach}. All experiments were repeated five times according to the experimental setup proposed by \cite{breakhis}, which is a five-fold repeated holdout. The proportion of data for training and test is 70\% and 30\%
patient-wise respectively, which means that no patient has parts of his/her images in training or test simultaneously. Another important aspect of the experiments is that we did not mix images of different magnifications. As a result of the experimental setup, we performed 200 executions.

%%%%%%%%%%%%%%%%%%%%%%%%%%%%%%%%%%%%%%%%%%%%%%%
\section{Results and Discussion}
\label{sec:results}
%%%%%%%%%%%%%%%%%%%%%%%%%%%%%%%%%%%%%%%%%%%%%%%
The experiment results are summarized in Table \ref{tab:accuracy}. Our results are reported following image-level and patient-level. For image-level, we also present the balanced accuracy, which in this two-class problem is the mean of specificity and sensitivity. All results are the mean of the five folds followed by their standard deviation. Our approach performed better than other methods compared with 400$\times$ and 200$\times$ magnifications. We can also highlight the result for Aug2$\times$ and 200$\times$ magnifications where the patient's accuracy is high, close to our method, but when looking at the balanced accuracy, it does not present good balanced image-level accuracy. In other words, it only performed well in one class, which shows the importance of presenting balanced accuracy due to the dataset imbalance. The approaches of mixing images, both by GLPB and Mix presented good results with a small advantage margin to GLPB.
\begin{table}[htbp!]
\caption{Five-fold mean image-level accuracy, image-level balanced accuracy and patient-level accuracy for four magnification factors. The best values are higlighted in bold font.}
\label{tab:accuracy}
\centering
\scriptsize
\setlength{\tabcolsep}{0.3em} % for the horizontal padding
{\renewcommand{\arraystretch}{1.2}% for the vertical padding
\begin{tabular}{|r r|r|r|r|r|}
\hline
 & & \multicolumn{4}{c|}{\bf Magnification}
\\
\cline{3-6}
 \multicolumn{2}{|c|}{\bf Method} & \multicolumn{1}{c|}{\bf 400$\times$} & \multicolumn{1}{c|}{\bf 200$\times$} & \multicolumn{1}{c|}{\bf 100$\times$} & \multicolumn{1}{c|}{\bf 40$\times$}\\
\hline
\parbox[t]{2mm}{\multirow{10}{*}{\rotatebox[origin=c]{90}{\bf Image}}} & GLPB     & \textbf{0.872 $\pm$ 0.045} & \textbf{0.884 $\pm$ 0.050} & 0.814 $\pm$ 0.048 & 0.821 $\pm$ 0.064 \\
& GLPB2$\times$     & 0.851 $\pm$ 0.061 & 0.884 $\pm$ 0.058 & \textbf{0.830 $\pm$ 0.057} & 0.808 $\pm$ 0.071 \\
& GLPB6$\times$     & 0.837 $\pm$ 0.037 & 0.876 $\pm$ 0.049 & 0.823 $\pm$ 0.046 & 0.787 $\pm$ 0.054 \\
& MixB\&M     & 0.871 $\pm$ 0.050 & 0.883 $\pm$ 0.050 & 0.824 $\pm$ 0.045 & \textbf{0.841 $\pm$ 0.055} \\
& MixSub & 0.821 $\pm$ 0.065 & 0.874 $\pm$ 0.059 & 0.822 $\pm$ 0.055 & 0.803 $\pm$ 0.065 \\
& NoAug       & 0.692 $\pm$ 0.067 & 0.816 $\pm$ 0.063 & 0.698 $\pm$ 0.026 & 0.678 $\pm$ 0.034 \\
& Aug2$\times$       & 0.798 $\pm$ 0.064 & 0.874 $\pm$ 0.040 & 0.771 $\pm$ 0.011 & 0.745 $\pm$ 0.063 \\
& GAN & 0.851 $\pm$ 0.050 & 0.870 $\pm$ 0.056 & 0.804 $\pm$ 0.042 & 0.774 $\pm$ 0.085 \\
& GAN2$\times$ & 0.849 $\pm$ 0.057 & 0.874 $\pm$ 0.057 & 0.821 $\pm$ 0.032 & 0.805 $\pm$ 0.067 \\
& GAN6$\times$ & 0.839 $\pm$ 0.041 & 0.877 $\pm$ 0.049 & 0.813 $\pm$ 0.021 & 0.795 $\pm$ 0.049 \\
\hline
\parbox[t]{2mm}{\multirow{10}{*}{\rotatebox[origin=c]{90}{\bf Balanced}}} & GLPB     & \textbf{0.847 $\pm$ 0.043} & 0.875 $\pm$ 0.063 & 0.795 $\pm$ 0.044 & 0.803 $\pm$ 0.049 \\
& GLPB2$\times$     & 0.845 $\pm$ 0.056 & \textbf{0.878 $\pm$ 0.068} & \textbf{0.821 $\pm$ 0.064} & \textbf{0.808 $\pm$ 0.069} \\
& GLPB6$\times$     & 0.831 $\pm$ 0.033 & 0.868 $\pm$ 0.062 & 0.820 $\pm$ 0.047 & 0.795 $\pm$ 0.051 \\
& MixB\&M     & 0.846 $\pm$ 0.051 & 0.866 $\pm$ 0.065 & 0.793 $\pm$ 0.064 & 0.807 $\pm$ 0.065 \\
& MixSub  & 0.818 $\pm$ 0.063 & 0.861 $\pm$ 0.069 & 0.793 $\pm$ 0.063 & 0.792 $\pm$ 0.074 \\
& NoAug       & 0.547 $\pm$ 0.073 & 0.720 $\pm$ 0.073 & 0.543 $\pm$ 0.030 & 0.513 $\pm$ 0.026 \\
& Aug2$\times$       & 0.711 $\pm$ 0.087 & 0.819 $\pm$ 0.053 & 0.661 $\pm$ 0.010 & 0.617 $\pm$ 0.095 \\
& GAN & 0.816 $\pm$ 0.057 & 0.847 $\pm$ 0.064 & 0.763 $\pm$ 0.043 & 0.709 $\pm$ 0.129 \\
& GAN2$\times$ & 0.834 $\pm$ 0.052 & 0.853 $\pm$ 0.063 & 0.788 $\pm$ 0.038 & 0.769 $\pm$ 0.061 \\
& GAN6$\times$ & 0.821 $\pm$ 0.033 & 0.855 $\pm$ 0.050 & 0.768 $\pm$ 0.026 & 0.761 $\pm$ 0.056 \\
\hline
\parbox[t]{2mm}{\multirow{10}{*}{\rotatebox[origin=c]{90}{\bf Patient}}} & GLPB & \textbf{0.882 $\pm$ 0.043} & \textbf{0.896 $\pm$ 0.054} & 0.835 $\pm$ 0.022 & 0.845 $\pm$ 0.042 \\
& GLPB2$\times$ & 0.857 $\pm$ 0.061 & 0.891 $\pm$ 0.067 & \textbf{0.852 $\pm$ 0.047} & 0.825 $\pm$ 0.067 \\
& GLPB6$\times$ & 0.847 $\pm$ 0.032 & 0.887 $\pm$ 0.055 & 0.841 $\pm$ 0.048 & 0.789 $\pm$ 0.063 \\
& MixB\&M     & 0.877 $\pm$ 0.053 & 0.890 $\pm$ 0.056 & 0.835 $\pm$ 0.029 & \textbf{0.860 $\pm$ 0.049} \\
& MixSub  & 0.823 $\pm$ 0.053 & 0.885 $\pm$ 0.063 & 0.834 $\pm$ 0.044 & 0.826 $\pm$ 0.059 \\
& NoAug       & 0.711 $\pm$ 0.054 & 0.830 $\pm$ 0.053 & 0.708 $\pm$ 0.025 & 0.689 $\pm$ 0.022 \\
& Aug2$\times$       & 0.815 $\pm$ 0.066 & 0.891 $\pm$ 0.035 & 0.771 $\pm$ 0.021 & 0.755 $\pm$ 0.066 \\
& GAN 	& 0.856 $\pm$ 0.053 & 0.884 $\pm$ 0.056 & 0.813 $\pm$ 0.046 & 0.796 $\pm$ 0.073 \\
& GAN2$\times$	& 0.857 $\pm$ 0.062 & 0.883 $\pm$ 0.062 & 0.842 $\pm$ 0.025 & 0.825 $\pm$ 0.057 \\
& GAN6$\times$	& 0.850 $\pm$ 0.047 & 0.886 $\pm$ 0.055 & 0.835 $\pm$ 0.023 & 0.814 $\pm$ 0.043 \\
\hline
\end{tabular}
}
\end{table}
Remarkably, 200$\times$ magnification is the most significant result for our proposal, presenting the best accuracy (image, image balanced, and patient accuracy). Table~\ref{tab:stateart} shows that the results achieved by the GLPB outperform some state-of-the-art results. We included in Table~\ref{tab:stateart} only results that followed the same fold distribution. Additional results on the BreakHis dataset using different fold distributions can be found in~\cite{reviewbreakhis}. In Table~\ref{tab:stateart}, the Baseline method is based on handcrafted feature extractors for texture and obtained better results with 200$\times$ magnification. AlexNet and Deep Features methods are based on CNNs and present better results in 100$\times$ and 40$\times$ magnifications. Our approach used a compact CNN, designed for texture recognition, which also presented good results for 200$\times$ magnification.

\begin{figure}
     \centering
     \begin{subfigure}[b]{0.23\textwidth}
         \centering
         \includegraphics[width=3.5cm]{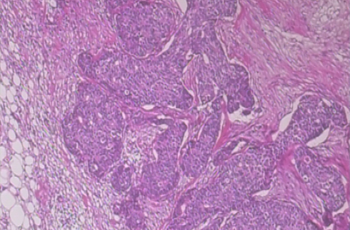}
         \caption{DC 40$\times$}
         \label{fig:ductal40}
     \end{subfigure}
     \hfill
     \begin{subfigure}[b]{0.23\textwidth}
         \centering
         \includegraphics[width=3.5cm]{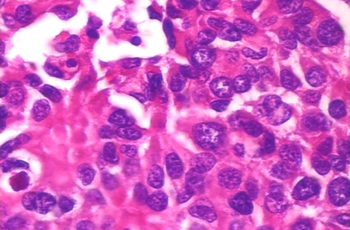}
         \caption{DC 400$\times$}
         \label{fig:ductal400}
     \end{subfigure}
     \\
     \begin{subfigure}[b]{0.23\textwidth}
         \centering
         \includegraphics[width=3.5cm]{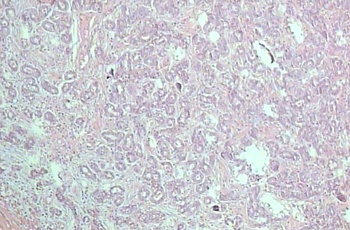}
         \caption{A 40$\times$}
         \label{fig:adenoma40}
     \end{subfigure}
     \hfill
     \begin{subfigure}[b]{0.23\textwidth}
         \centering
         \includegraphics[width=3.5cm]{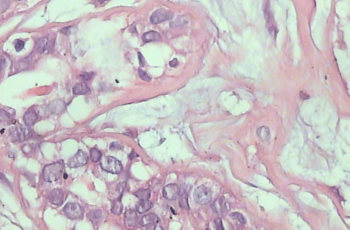}
         \caption{A 400$\times$}
         \label{fig:adenoma400}
     \end{subfigure}
        \caption{Comparison of Ductal Carcinoma (DC) and Adenoma (A) images with magnifications of (a) 40$\times$  (b) 400$\times$ (c) 40$\times$ (d) 400$\times$ }
        \label{fig:magnificationimages}
\end{figure}

We present in Fig.~\ref{fig:magnificationimages} an example of HIs of two tumor types with two different magnifications. Figs.~\ref{fig:magnificationimages}a and \ref{fig:magnificationimages}b are Ductal Carcinoma images of 40$\times$ and 400$\times$ magnifications. Figs.~\ref{fig:magnificationimages}a and \ref{fig:magnificationimages}b are Adenoma images of 40$\times$  and 400$\times$ magnification factors. These images highlight the differences in the objective into which the filters are trained. In the 40$\times$ magnification, it is difficult to identify nucleus pleomorphism since they are too small. On the other hand, this magnification makes easy to detect forms like the ones that characterize papillary carcinoma. The characteristics of magnifications impact in the network used, taking into account that the small network does not have the capabilities of large objects recognizing, only textures.
To the same extent, future work studies may include shallow methods in order to compare the performance of GLPB vis-à-vis other techniques of undersampling and oversampling such as SMOTE and NearMiss algorithms.

\begin{table}[htb]
\caption{State-of-the-art patient accuracy results with the same fold split.}
\label{tab:stateart}
\centering
\scriptsize
\setlength{\tabcolsep}{0.3em} % for the horizontal padding
{\renewcommand{\arraystretch}{1.2}% for the vertical padding
\begin{tabular}{|c|r|r|r|r|}
\hline
& \multicolumn{4}{c|}{\bf Magnification} \\
\cline{2-5}
\bf Method & \multicolumn{1}{c|}{\bf 400$\times$} & \multicolumn{1}{c|}{\bf 200$\times$} & \multicolumn{1}{c|}{\bf 100$\times$} & \multicolumn{1}{c|}{\bf 40$\times$} \\
\hline
Baseline \cite{breakhis} & 0.823 $\pm$ 0.038 & 0.851 $\pm$ 0.031 & 0.821 $\pm$ 0.049 & 0.838 $\pm$ 0.041 \\
AlexNet \cite{alexbreakhis} & 0.817 $\pm$ 0.049 & 0.853 $\pm$ 0.038 & 0.845 $\pm$ 0.024 & 0.886 $\pm$ 0.056 \\
Deep Features \cite{deepfeatures} & 0.861 $\pm$ 0.062 & 0.863 $\pm$ 0.035 & 0.884 $\pm$ 0.048 & 0.840 $\pm$ 0.069 \\ 
Mi \cite{milearning} &  0.827 $\pm$ 0.030 & 0.872 $\pm$ 0.043 & \bfseries 0.891 $\pm$ 0.052 & \bfseries 0.921 $\pm$ 0.059 \\
GLPB & \bfseries 0.882 $\pm$ 0.043 & \bfseries 0.896 $\pm$ 0.054 & 0.835 $\pm$ 0.022 & 0.845 $\pm$ 0.042 \\
\hline
\end{tabular}
}
\end{table}

Finally, Table~\ref{tab:subclass40x} presents the subclass classification result for 40$\times$ and 200$\times$ magnifications using GLPB2$\times$. These are the two scenarios where GLPB achieved the worst and the best performance, respectively. We consider 40$\times$ magnifications the worst result because the difference to the state-of-the-art is the highest one. We also show the Aug2$\times$ result for 200$\times$ magnification that achieved good patient accuracy results but presented poor balanced accuracy results. It can be seen in Table~\ref{tab:subclass40x}, where the Malignant tumors were much better identified than the Benign.

\begin{table}[htb]
\caption{Classification accuracy considering subclass distribution for 40$\times$ and 200$\times$ magnifications using GLPB2$\times$ and 200$\times$ magnification with Aug2$\times$. Adenoma (A), Fibroadenoma (F), Tubular Adenoma (TA), Phyllodes Tumor (PT), Ductal Carcinoma (DC), Papillary Carcinoma (PC), Mucinous Carcinoma (MC) and Lobular Carcinoma (LC)}
\label{tab:subclass40x}
\centering
\scriptsize
\setlength{\tabcolsep}{0.4em} % for the horizontal padding
{\renewcommand{\arraystretch}{1.2}% for the vertical padding
\begin{tabular}{|ll|rr|rr|rr|rr|rr|}
\hline
& & \multicolumn{10}{c|}{\bf Predicted} \\
\cline{3-12}
& & \multicolumn{1}{c}{\bf B} & \multicolumn{1}{c|}{\bf M} & \multicolumn{1}{c}{\bf B} & \multicolumn{1}{c|}{\bf M} & \multicolumn{1}{c}{\bf B} & \multicolumn{1}{c|}{\bf M} & \multicolumn{1}{c}{\bf B} & \multicolumn{1}{c|}{\bf M} &
\multicolumn{1}{c}{\bf B} & \multicolumn{1}{c|}{\bf M} \\
\hline
& & \multicolumn{10}{c|}{\bf Mag 40$\times$ GLPB2$\times$} \\
\cline{3-12}
\parbox[t]{3mm}{\multirow{4}{*}{\rotatebox[origin=c]{90}{\bf Benign}}} & \bf A & 0.71 & 0.29 & 0.68 & 0.32 & 1.00 & 0.00 & 1.00 & 0.00 & 0.88 & 0.13 \\
& \bf F  & 0.91 & 0.09 & 0.85 & 0.15 & 0.77 & 0.23 & 0.75 & 0.25 & 0.78 & 0.22 \\
& \bf TA & 0.73 & 0.27 & 0.56 & 0.44 & 1.00 & 0.00 & 0.48 & 0.52 & 0.51 & 0.49 \\
& \bf PT & 0.74 & 0.26 & 1.00 & 0.00 & 1.00 & 0.00 & 1.00 & 0.00 & 0.55 & 0.45 \\
\hline
\parbox[t]{3mm}{\multirow{4}{*}{\rotatebox[origin=c]{90}{\bf Malignant}}} & \bf DC & 0.12 & 0.88 & 0.08 & 0.92 & 0.10 & 0.90 & 0.23 & 0.77 & 0.05 & 0.95 \\
& \bf PC & 0.62 & 0.38 & 0.75 & 0.25 & 0.11 & 0.89 & 0.69 & 0.31 & 0.18 & 0.82 \\
& \bf MC & 0.13 & 0.87 & 0.40 & 0.60 & 0.11 & 0.89 & 0.44 & 0.56 & 0.07 & 0.93 \\
& \bf LC & 0.11 & 0.89 & 0.07 & 0.93 & 0.00 & 1.00 & 0.04 & 0.96 & 0.00 & 1.00 \\
\hline
& & \multicolumn{10}{c|}{\bf Mag 200$\times$ GLPB2$\times$} \\
\cline{3-12}
\parbox[t]{3mm}{\multirow{4}{*}{\rotatebox[origin=c]{90}{\bf Benign}}} & \bf A  & 0.81 & 0.19 & 0.56 & 0.44 & 0.98 & 0.02 & 0.98 & 0.02 & 0.96 & 0.04 \\
&\bf F  & 0.98 & 0.02 & 1.00 & 0.00 & 1.00 & 0.00 & 0.72 & 0.28 & 0.97 & 0.03 \\
&\bf TA & 0.86 & 0.14 & 0.55 & 0.45 & 1.00 & 0.00 & 0.58 & 0.42 & 0.90 & 0.10 \\
&\bf PT & 0.75 & 0.25 & 0.93 & 0.07 & 0.93 & 0.07 & 1.00 & 0.00 & 0.78 & 0.22 \\
\hline
\parbox[t]{3mm}{\multirow{4}{*}{\rotatebox[origin=c]{90}{\bf Malignant}}} & \bf DC & 0.09 & 0.91 & 0.05 & 0.95 & 0.07 & 0.93 & 0.11 & 0.89 & 0.05 & 0.95 \\
& \bf PC & 0.32 & 0.68 & 0.69 & 0.31 & 0.00 & 1.00 & 0.47 & 0.53 & 0.02 & 0.98 \\
& \bf MC & 0.05 & 0.95 & 0.15 & 0.85 & 0.02 & 0.98 & 0.42 & 0.58 & 0.12 & 0.88 \\
& \bf LC & 0.09 & 0.91 & 0.00 & 1.00 & 0.00 & 1.00 & 0.00 & 1.00 & 0.00 & 1.00 \\
\hline
& & \multicolumn{10}{c|}{\bf Mag 200$\times$ Aug2$\times$} \\
\cline{3-12}
\parbox[t]{3mm}{\multirow{4}{*}{\rotatebox[origin=c]{90}{\bf Benign}}} & \bf A & 0.40 & 0.60 & 0.42 & 0.58 & 0.66 & 0.34 & 0.65 & 0.35 & 0.63 & 0.37 \\
& \bf F  & 0.82 & 0.18 & 1.00 & 0.00 & 0.98 & 0.02 & 0.71 & 0.29 & 0.66 & 0.34 \\
& \bf TA & 0.49 & 0.51 & 0.55 & 0.45 & 0.90 & 0.10 & 0.51 & 0.49 & 0.46 & 0.54 \\
& \bf PT & 0.54 & 0.46 & 0.79 & 0.21 & 0.71 & 0.29 & 0.76 & 0.24 & 0.54 & 0.46 \\
\hline
\parbox[t]{3mm}{\multirow{4}{*}{\rotatebox[origin=c]{90}{\bf Malignant}}} & \bf DC & 0.01 & 0.99 & 0.02 & 0.98 & 0.01 & 0.99 & 0.01 & 0.99 & 0.00 & 1.00 \\
& \bf PC & 0.11 & 0.89 & 0.54 & 0.46 & 0.00 & 1.00 & 0.06 & 0.94 & 0.00 & 1.00 \\
& \bf MC & 0.00 & 1.00 & 0.04 & 0.96 & 0.00 & 1.00 & 0.04 & 0.96 & 0.01 & 0.99 \\
& \bf LC & 0.01 & 0.99 & 0.00 & 1.00 & 0.00 & 1.00 & 0.00 & 1.00 & 0.00 & 1.00 \\
\hline
\end{tabular}
}
\end{table}

%%%%%%%%%%%%%%%%%%%%%%%%%%%%%%%%%%%%%%%%%%%%%%%
\section{Conclusion}
\label{sec:conclusion}
%%%%%%%%%%%%%%%%%%%%%%%%%%%%%%%%%%%%%%%%%%%%%%%
%The work herein presents a noteworthy methodology regarding the use of Gaussian and Laplacian pyramids methods concerning DA.
This paper presented a novel method for data augmentation based on Gaussian and Laplacian pyramids. We investigated how the classifiers can be improved in terms of the generalization ability considering the inter-patient variability. We stated that a DA could be decomposed into image blending such that from two images the third can be generated, where each one is from a different patient while producing both a balanced dataset and good classification accuracy, precision, and recall. 

It is essential to consider that Gaussian-Laplacian pyramids provide a powerful mathematical tool for representing multi-resolutions of an image as well as a blending between two images of the same class while finding the most favorable seam and avoiding artifacts along boundaries. Thence, in this paper, blending images helped to augment the dataset and to balance the respective classes under investigation, meanwhile performing at comparable accuracy as other state-of-the-art methods. The blending method can provide improvements on the results of the other state-of-the-art methods, considering that it works as a preprocessing stage and can be used together
%allied
with
%the reported literature best
other methods as well. Accordingly, the proposed method is promising to circumvent the limitation of data, in particular for deep methods where the classification considers HIs as whole images.
%, and not as using patches}. 

\balance
\bibliographystyle{IEEEbib}

\end{document}